# CHSH inequality test via disturbance-free measurement

Sang Min Lee[1,2],[*] Minsu Kim[1], Heonoh Kim[1], Han Seb Moon[1],[†] and Sang Wook Kim[3][‡]

[1]*Department of Physics, Pusan National University, Busan 46241, South Korea*
[2]*Korea Research Institute of Standards and Science, Daejeon 34113, South Korea and*
[3]*Department of Physics Education, Pusan National University, Busan 46241, South Korea*

(Dated: June 30, 2018)

## Abstract

We propose disturbance-free measurement using a "weak-value" scheme, in which a weakly measured quantum system is post-selected (to the initial state) to confirm that there is no disturbance. The probability of obtaining the non-disturbed state is asymptotically close to unity. We theoretically show that outcomes of the disturbance-free measurement for a two-qubit state satisfy the Clauser–Horne–Shimony–Holt inequality. We experimentally demonstrate the test for a typical (maximally entangled) two-qubit state based on a linear optical system. In experiments, polarization-entangled photon-pairs generated by the spontaneous parametric down-conversion process are measured by instruments such as strength-variable polarization-measurement apparatuses and a fiber-based Bell state analyzer.

---

[*] samini@kriss.re.kr

[†] moon.hanseb@gmail.com

[‡] swkim0412@gmail.com

## I. INTRODUCTION

One of the strange features of quantum mechanics is the inevitable measurement-induced disturbance (MID), which manifests in the original description of Heisenberg's uncertainty principle [1–3], although its interpretation remains under debate [4, 5]. In quantum mechanics, measurement and information gathering disturb and alter the state of the measured system [6–12]. In contrast, in classical mechanics, one can obtain information from a system without disturbing it. The measurement outcome directly reflects the property that the observed system has already owned before measurement, which guarantees the "realism" of classical mechanics. Combining the "locality," local-realism was believed to be one of the fundamental bases of classical physics and led to a debate regarding the Einstein–Podolsky–Rosen (EPR) paradox [13]. Bell's inequality is designed to test whether a physical state satisfies local-realism from the viewpoint of hidden-variable theory [14]. Thus far, many tests of Bell's inequality have been performed experimentally [15–17], revealing that local-realism is false. Now, we know that quantum states cannot be fully described by classical theory. However, it is still unclear what the classicality really means and how it can emerge from the quantumness. It is quantum coherence that usually explains the quantum-to-classical transition [18]; a quantum state loses its quantum coherence (or correlation) owing to the interaction with the external environment. Recently, it has been found that imprecise or inaccurate measurements [19, 20] also lead to the quantum-to-classical transition.

Measurement inevitably induces disturbance to the measured system in quantum mechanics, whereas it does not in classical mechanics. Thus, we may assume that the disturbance induced by measurement could be a criterion for the distinction between classicality and quantumness. In this paper, we pose a question: *"If we gather information from a quantum system without disturbing it, does the outcome exhibit classical behavior?"* To address this question, we propose an approximate disturbance-free measurement (DFM) in quantum mechanics using a weak-value scheme and discuss the meaning of the classicality in the context of Bell's inequality. To obtain the correlations of Bell's inequality in the most general setting, we calculate the expectation values of joint observables observed with an arbitrary measurement-strength. We show that a quantum state satisfies Bell's inequality if it is measured without disturbance. We perform a proof-of-principle experiment using a maximally entangled state (MES), which is not only the most non-classical state but also exhibits the most dramatic change via DFM. In section II, we present our main idea and theory. In section III, we discuss the experimental setup, which consists of a polarization-entangled photon-pair source, strength-variable polarization-measurement apparatuses, and a fiber-based Bell-state analyzer. In particular, we describe how to adjust the measurement-strength and align the apparatus and explain why the fiber-based analyzer is used. In section IV, we discuss the experimental results. Finally, we summarize and conclude the paper in section V. Detailed mathematical derivations and supplementary physical arguments are presented in the appendices.

## II. THEORY

### A. Disturbance-free measurement

MID alters the state of the measured system from an initial pure state $|\psi_i\rangle$ to a final density matrix $\rho$. We often define the infidelity, which quantifies the degree of disturbance of $|\psi_i\rangle$, as [6]

$$D \equiv 1 - \langle\psi_i|\rho|\psi_i\rangle. \tag{1}$$

The amount of disturbance decreases when the measurement-strength, denoted as $s$, decreases (see Appendix A). Within the weak-measurement limit ($0 < s \ll 1$), the disturbance is negligible ($D \ll 1$). Therefore, one naive approach for obtaining information from a quantum system without disturbance is to measure the system very weakly. However, regardless of the degree of measurement "weakness," the measured state is disturbed ($D \neq 0$) [12]. Thus, there is always a non-zero probability that the measured state is orthogonal to the original state. Thus, the weak-measurement condition is insufficient for achieving DFM.

Our approach to realizing DFM is as follows (see also Figure 1). First, similar to the above naive approach, we very weakly measure an observable $M$ of a system in the state $|\psi_i\rangle$. We obtain the outcome $m$, and the state is changed to $\rho$. In general, $|\psi_i\rangle$ is not an eigenstate of $M$. We then perform a projection measurement on $\rho$. If and only if $\rho$ is collapsed to $|\psi_i\rangle$, the outcomes of the weak measurement are selected and labeled as $m^*$. If the measurement is sufficiently weak, the majority of the outcomes are selected as $|\psi_i\rangle$, because $\rho$ does not differ significantly from $|\psi_i\rangle$. This is guaranteed because the disturbance $D$ is close to zero in the weak-measurement limit (see Appendix A). Thus, we can state with confidence that each selected outcome $m^*$ is information obtained without disturbing the system, as $m^*$ is only recorded if we confirm the nondisturbance of the state. Simultaneously, the probability of obtaining $|\psi_i\rangle$ in the final projection measurement approaches 1. The average of such conditional outcomes of weak measurement over the post-selected data is called the *weak-value*, $\langle M \rangle_w$ [21]. Thus, the outcome of DFM is similar with the weak-value in the condition that the post-selected state is equivalent to the initial state, implying that the initial state in some sense is not disturbed.

Note that if we average over all $m$, ignoring the results of the final projection measurement (post-selections), we simply obtain the quantum-mechanical average $\langle M \rangle$. This is because the distribution of $m$ for the weak measurement is Gaussian with a large width due to the weak measurement but with the average equivalent to the quantum-mechanical average (see Appendix B). We also note that the outcome of DFM for a single observable is exactly equivalent to the ordinary quantum mechanical average $\langle M \rangle$, as $\langle M \rangle_{\text{DFM}} = \langle M \rangle_{w,(i=f)} = \langle \psi_f | M | \psi_i \rangle / \langle \psi_f | \psi_i \rangle = \langle \psi_i | M | \psi_i \rangle = \langle M \rangle$. However, the DFM outcome of join observables, i.e., $A$ and $B$, measured by separable interactions and pointer states does not have to be the same as the similar quantum-mechanical average, i.e., $\langle AB \rangle_{\text{DFM}} \neq \langle AB \rangle_{w,(i=f)} = \langle AB \rangle$, but is induced as,

$$\langle \hat{A}\hat{B} \rangle_{\text{DFM}} = \frac{1}{2}\left( \langle \hat{A}\hat{B} \rangle + \langle \hat{A} \rangle \langle \hat{B} \rangle \right). \tag{2}$$

As described in Appendix C, we define $\langle AB \rangle_{\text{DFM}}$ as the measurement outcome of local single-particle interactions in the weak limit [22] when the system is post-selected to the initial state, and analytically derive expectation values of joint observables for an arbitrary measurement-strength and any post-selections, which are expanded from previous works for a single observable [23–26]. We note two points as follows. The expectation value of any observable is obtained from the average value of the pointer state (see Appendix B), and for the joint observables of CHSH inequality independent measurement on each site is a natural assumption from the view-point of the locality (see Appendix C).

### B. Identification of classical behavior via CHSH inequality

Numerous studies have examined approaches for obtaining classicality from quantumness [19, 20, 27, 28]. Here, as in previous papers, we exploit the Clauser–Horne–Shimony–Holt (CHSH) inequality [29] to determine whether the outcomes behave classically [30]. The

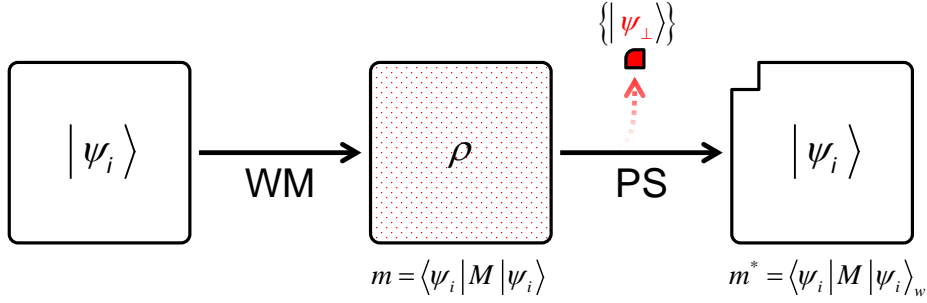

FIG. 1. Schematic of DFM. An observable $M$ of the initial state $|\psi_i\rangle$ is very weakly measured, so as to be infinitesimally modified to $\rho$. The average of the measured outputs is equivalent to the quantum-mechanical average, $m = \langle\psi_i|M|\psi_i\rangle$. DFM is achieved when the average of only the $M$ outcomes post-selected onto the initial state $|\psi_i\rangle$ is considered: $m^* = \langle\psi_i|M|\psi_i\rangle_w$; this simply corresponds to one specific case of the weak-value scheme. The outcomes of the post-selection are mostly $|\psi_i\rangle$ with a tiny portion of $\{|\psi_\perp\rangle\}$, i.e., the subset orthogonal to $|\psi_i\rangle$. WM: weak measurement; PS: post-selection.

CHSH inequality reads

$$|S| = |C(A, B) + C(A, B') + C(A', B) - C(A', B')| \leq 2, \qquad (3)$$

where $C$ denotes the correlation between two observables $A$ (or $A'$) and $B$ (or $B'$). With the Bell state, $|S|$ maximally violates this inequality to reach $2\sqrt{2}$, and such an inequality violation unambiguously exhibits quantumness. To theoretically calculate the $S$ measured by DFM, $S_{\text{DFM}}$, for an arbitrary pure state $|\psi\rangle$, we use $\langle AB\rangle_{\text{DFM}}$ defined in equation (2), which is obtained from the joint expectation value of two pointers when the system is post-selected to the initial state, and is equivalent to the average of the correlation value, $\langle AB\rangle$, and the product of the expectation values of each observable, $\langle A\rangle\langle B\rangle$. Using equation (2), the $S$ of the CHSH inequality under DFM is expressed as

$$\begin{aligned}S_{\text{DFM}} &= \langle\hat{A}\hat{B}\rangle_{\text{DFM}} + \langle\hat{A}'\hat{B}\rangle_{\text{DFM}} + \langle\hat{A}\hat{B}'\rangle_{\text{DFM}} - \langle\hat{A}'\hat{B}'\rangle_{\text{DFM}}, \\ &= \frac{1}{2}(S_I + S_{II}),\end{aligned} \qquad (4)$$

where $S_I = \langle\hat{A}\hat{B}\rangle + \langle\hat{A}'\hat{B}\rangle + \langle\hat{A}\hat{B}'\rangle - \langle\hat{A}'\hat{B}'\rangle$ and $S_{II} = \langle\hat{A}\rangle\langle\hat{B}\rangle + \langle\hat{A}'\rangle\langle\hat{B}\rangle + \langle\hat{A}\rangle\langle\hat{B}'\rangle - \langle\hat{A}'\rangle\langle\hat{B}'\rangle$.

We expect that the maximum value of $S_{\text{DFM}}$, $S_{\text{DFM}}^{\max}$, for an arbitrary pure state is a function of the degree of entanglement. We easily estimate the $S_{\text{DFM}}^{\max}$ in two extreme cases: the MES and a separable state. For the MES, we find the maximum value of $\sqrt{2}$, as $S_I = 2\sqrt{2}$ and $S_{II} = 0$, because the expectation value of any single observable for MES is 0: $\langle \hat{A} \otimes \hat{I} \rangle_{\text{MES}} = 0$. For separable states, we obtain the maximum value of 2, as $S_I = S_{II} = 2$, because $\langle \hat{A}\hat{B} \rangle = \langle \hat{A} \rangle \langle \hat{B} \rangle$. Figure 2 presents the numerically calculated $S_{\text{DFM}}^{\max}$ as a function of the concurrence [31] for all possible pure states. This figure clearly shows that the upper bound of $S_{\text{DFM}}$ for arbitrary states is equal to 2. Thus, our main query is answered in the affirmative, i.e., if we gather information from a quantum system without disturbance (via DFM), the outcome is limited to the classical bound ($|S_{\text{DFM}}| \leq 2$). As shown in Figure 2, the maximum value of $S$ is reduced most dramatically for MES; thus, our experimental tests were performed on an MES.

Note also that (a) the result of CHSH test measured by our DFM can be described by the local hidden variable model (see Appendix D); and (b) with only a single observable at different times, one can test the Leggett–Garg inequality based on macroscopic realism [32]. However, we demonstrate that a trivial result is obtained (see Appendix E).

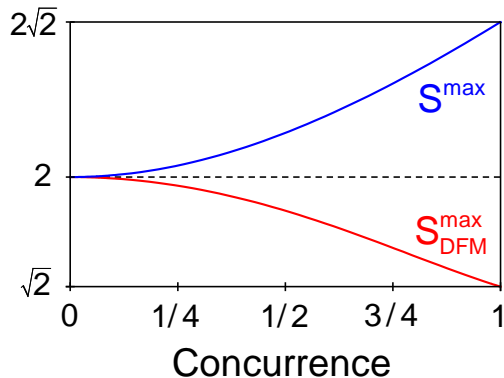

FIG. 2. $S_{\text{DFM}}^{\max}$ (in red): Maximum values of $S$ measured by DFM are obtained from numerical calculations as a function of the concurrence. $S^{\max}$ (in blue): the maximum value of the ordinary CHSH inequality is also represented, for comparison.

## III. EXPERIMENTS

Before describing our experimental setup, we note the following. In real experiments, it is difficult to realize the weak-measurement limit $s \to 0$, because at this limit we need an infinite number of measurement trials to obtain a meaningful value of the measurand (expectation value with relatively small uncertainty). Thus, we perform experiments (measure correlations) as a function of $s$, to verify the correspondence between theory and experiment and finally to obtain $S_{\text{DFM}}$ asymptotically by decreasing $s$. We expect that $|S|$, beginning with a value larger than the classical bound 2, decreases and moves below this bound as $s \to 0$. For both limits, i.e., $s \to \infty$ and $s \to 0$, $|S|$ can be calculated using our theory in Appendix C. In addition, because the initial state in the experiment is an MES, a special experiment setup is needed to simultaneously perform the post-selection to the initial state and measure the distribution of the pointer (measurement device) state. Thus, we develop a single-mode fiber (SMF)-based Bell-state analyzer.

### A. Experimental setup

Figure 3 shows our experimental setup, which consists of three parts.

#### 1. Pre-selection of $|\psi^-\rangle$: SMF-coupled polarization-entangled photon-pair source.

Using a spontaneous parametric down-conversion (SPDC) process of a periodically poled KTiOPO$_4$ (PPKTP) crystal ($\Lambda$=10 $\mu$m, $L$=10 mm) and a single-longitudinal-mode continuous-wave (cw) pump laser ($\lambda \simeq 406$ nm, $\Delta\nu = 0.2(1)$ GHz), we generate wavelength-degenerated ($\simeq 812$ nm) SMF-coupled polarization-entangled photon pairs [33]. The state before coupling to SMF is one of triplet state as $|\phi^+\rangle = (|HH\rangle + |VV\rangle)/\sqrt{2}$. Here, $|H\rangle$ and $|V\rangle$ denote single-photon states linearly polarized in the horizontal and vertical directions, respectively. By employing fiber polarization controllers (PCs) and a polarization-dependent phase shifter $\phi_1$ (consisting of three wave-plates), we prepare the singlet state at the output of the SMFs, which is described as

$$|\psi^-\rangle = \frac{|HV\rangle - |VH\rangle}{\sqrt{2}}. \tag{5}$$

For a 4-mW pump power, the single counts of each port for $|H\rangle$ or $|V\rangle$ are approximately 180 kHz, and the net coincidence counts for $|HH\rangle$ and $|VV\rangle$ are 19 kHz. The experimentally measured fidelity and the polarization correlation visibilities [34] are larger than 95 % [33]. The spatial modes of the photons emitted from the SMFs are approximately Gaussian, which is a basic assumption of weak measurement [21].

### 2. Weak measurements: strength-variable polarization-measurement apparatus

Each photon of a photon pair traverses a strength-variable polarization measurement apparatus [35]. Depending on the $H$ and $V$ polarizations, the spatial mode of the input beam is split and recombined by beam displacers (BDs). According to Snell's law, two tilted glasses (2.5-mm thick) in both arms yield a polarization-dependent beam displacement at the output. The measurement-strength or displacement can be adjusted by tilting the glass angles, because the strength is defined as $s = g^2/(2\sigma^2)$ [23], where $g$ is half of the distance between the two beams, and $\sigma$ is the fixed width of the Gaussian intensity profile from the SMF. Thus, this device is a kind of polarization-based interferometer if not $s \gg 1$. The weak-measurement bases equivalent to the displacement polarization bases $(H, V)$ are effectively controlled by the half-wave plate (HWP) angle $\theta_i$.

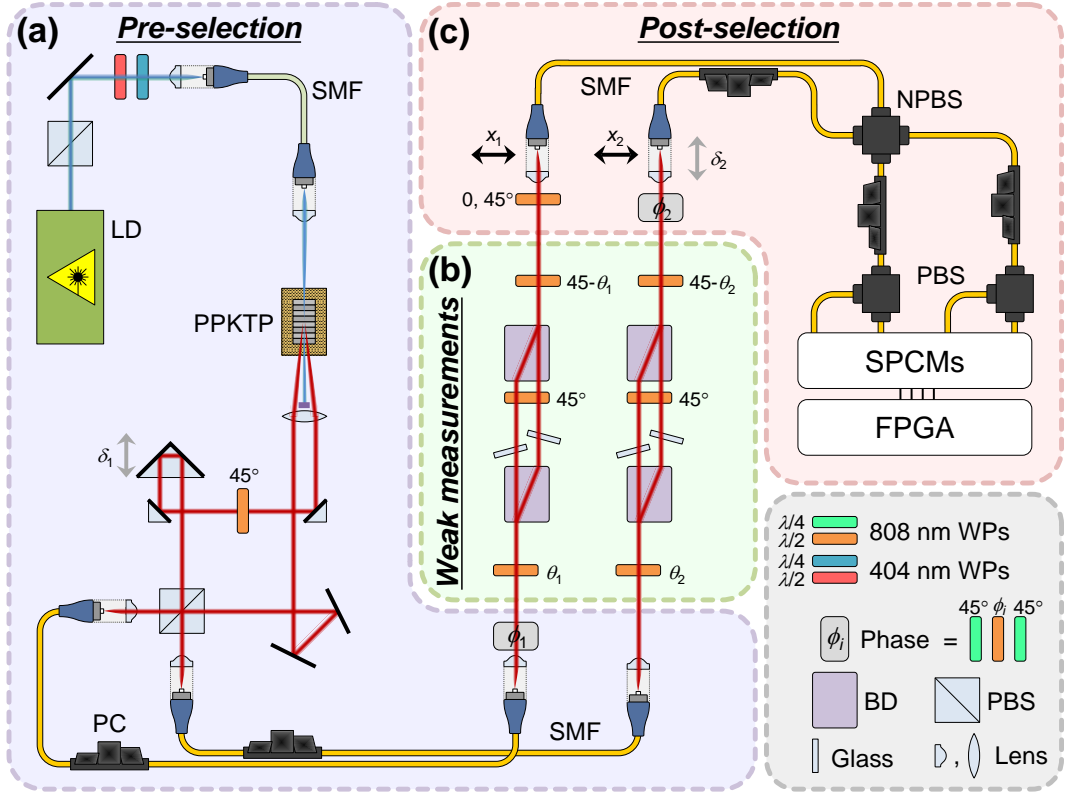

FIG. 3. Experimental setup. (a) Pre-selection. We use a non-collinear SPDC process with a PPKTP crystal and a compensation setup to generate a maximally polarization-entangled state. The polarization singlet state is prepared using PCs and $\phi_1$ to adjust the polarizations and phase in the SMFs. (b) Weak measurements. The spatial mode of each incidence beam is slightly separated into two parts. Their effective measurement bases are linearly polarized along $2\theta_i$ and $\frac{\pi}{2}+2\theta_i$ from the horizontal axis so as to be orthogonal to each other. (c) Post-selection. The fiber-based Bell-state analyzer consists of an NPBS, two PBSs, and coincidence counters. By adding an HWP (0° or 45°), we can post-select two states ($|\psi^{\pm}\rangle$ or $|\phi^{\pm}\rangle$) among the four Bell states. LD: laser diode at 406 nm, PPKTP: periodic-poled KTiOPO$_4$ crystal; $\delta_i$: optical delay line; PC: polarization controller; SMF: single-mode fiber; $\phi_i$: phase controller consisting of WPs; NPBS: non-polarizing beam splitter; PBS: polarizing beam splitter; SPCM: single photon counting module; FPGA: coincidence counter; WP: wave-plate; BD: calcite beam displacer.

When the glasses are tilted to control $s$, they induce an optical path length difference between the two arms. To remove this difference, we employ auxiliary setups (not shown in Figure 3). Two linear polarizers of $\pm 45°$ are temporally installed between the BDs and HWPs (denoted as $\theta_i$ or $45 - \theta_i$ in Figure 3). Using an SMF-coupled cw laser (at 810 nm) and a charge-coupled device camera, we measure the output-beam interference pattern. By finding the glass tilting angles that yield maximum visibility and centering the destructive interference patterns, we remove the optical path length differences. The phase between the two arms is stable for a period longer than three days.

3. *Post-selection, SMF-based Bell-state analyzer*

The weakly measured (or disturbed) singlet state is post-selected to two ($|\psi^\pm\rangle = |HV\rangle \pm |VH\rangle$ or $|\phi^\pm\rangle = |HH\rangle \pm |VV\rangle$) Bell states via an HWP (at 0° or 45°) and an SMF-based Bell-state analyzer [36], consisting of fiber-optics beam splitters, compensators (PCs, $\phi_2$ and $\delta_2$) and a coincidence counter. The coincidence count distributions (pointer-state distribution depending on post-selections) obtained by adjusting the fiber-coupler positions $x_i$ are employed to calculate the measurement outcomes, i.e., the expectation values $\langle AB \rangle$ or $\langle AB \rangle_w$ estimated from $\langle x_1 x_2 \rangle$ (see Appendices B and C). We note that the SMF-based Bell-state analyzer and the detection probability distribution measurements are the essential components of this experiment, because the conventional probability distribution measurement of the pointer state using a narrow slit [35, 37] is almost inapplicable to the Bell-state analyzer in free space. Generally, two beams passing through independently moving (scanning) slits do not meet at the non-polarizing beam splitter (NPBS) of the Bell-state analyzer.

### B. Advantages of SMF-based Bell-state analyzer

Our SMF-based Bell-state analyzer yields two advantages. First, it ensures that the spatial modes of the two input beams entering the NPBS in the SMF-based Bell-state analyzer are identical. Second, we can achieve an effectively weaker measurement strength. The scanning process of SMF-to-SMF coupling ensures that the Gaussian profiles of the input beams remain unchanged, with only the beam widths being increased. We can redefine the beam-profile width as an effective value, such that $\vec{E}_1\ G(c_1, \sigma_1) + \vec{E}_2\ G(c_2, \sigma_1) \xrightarrow{*G(\sigma_2)} \vec{E}_1\ G(c_1, \sigma_3) + \vec{E}_2\ G(c_2, \sigma_3)$, where the two terms headed by $\vec{E}_i$ are separated two beams via a strength-variable measurement apparatus, and $G(c_i, \sigma_j)$ is a Gaussian function with center position $c_i$ and width $\sigma_j$, and $\sigma_3^2 = \sigma_1^2 + \sigma_2^2$. In this Gaussian convolution process denoted as $*$, $\sigma_1$ is the original width of the initial beam (before weak measurement), $\sigma_2$ is the width of the scanning coupler (in the detection component), and $\sigma_3$ is the effective width of the measured profiles. $\vec{E}_i$ represents other degrees of freedom, such as the polarizations and electric-field amplitudes. With a fixed displacement $|c_2 - c_1|$ and an increased width ($\sigma_1 \to \sigma_3$), the strength of the weak measurement is effectively downscaled.

## IV. RESULTS AND DISCUSSIONS

### A. Experimentally measured pointer-state distributions and $\langle AB \rangle$

The quantum-mechanical counterparts of the classical polarization observables $A$ and $B$, i.e., the Stokes parameters, are Pauli operators defined as $\vec{n} \cdot \vec{\sigma}$, where $\vec{n}$ is a unit vector in the Bloch space and $\vec{\sigma}$ is the Pauli vector [38]. The maximum violation of the CHSH inequality in equation (2) occurs when the observables are chosen as $A = \hat{Z}$, $A' = \hat{X}$, $B = (\hat{X} + \hat{Z})/\sqrt{2}$, and $B' = (-\hat{X} + \hat{Z})/\sqrt{2}$. $\hat{X}$ ($\hat{Z}$) is a Pauli operator along the $x$ ($z$) axis in the Bloch space. The corresponding measurement bases are provided by respective HWP angles of $\theta_1 = 0$, $\theta_1' = \pi/8$, $\theta_2 = \pi/16$, and $\theta_2' = -\pi/16$ [34]. With the bases for the maximum violation of the CHSH inequality, we obtained detection probability (pointer state) distributions depending on four post-selected Bell states and calculated $\langle AB \rangle$ and $|S|$ using the expectation values of the pointer states. Figure 4 shows an example (case of $s \simeq 1$) of experimentally measured coincidence counting distributions with post-selection to the Bell states ($\psi^\pm, \phi^\pm$) or without post-selection ($\Sigma$, sum of all distributions). The corresponding theoretically calculated probability distributions of pointer states are shown in Appendix F. The $\langle AB \rangle$ and $|S|$, estimated by the $\langle x_1 x_2 \rangle$ of each distribution and their combination of four cases are shown at the top of each graph and at the bottom of each column, respectively. We normalize $x_i$ with $g_i$, i.e., $x_i/g_i$. The error bar (in pink) associated with each data point in Figure 4 represents the Poissonian photon-counting error ($\simeq \sqrt{N}$), which is relatively small. The experimental results are well-matched with the theoretical calculations shown in Figure 7.

### B. Comparison between measured and calculated $|S|$

We repeat the experiment with $s$ varying from 0.21 to 4.14, and summarize several $|S|$ values in Figure 5. Using the measurement outcomes, we obtain the $|S|$ for four Bell states as a function of $s$, denoted as $|S_{\psi^-}|$, $|S_{\psi^+}|$, $|S_{\phi^-}|$, and $|S_{\phi^+}|$, which are presented in Figures 5(a)–(d). As the DFM is achieved with $|\psi_i\rangle = |\psi_f\rangle$, our main interest lies in Figure 5(a), but the results with other post-selected states and without post-selection are shown in Figures 5(b)–(e) to show the consistency between the experiments and theoretical calculations. The theoretical expectations (red curves) fit the experimental results (dots) reasonably well.

### C. Experimentally obtained approximate value of $|S_{\text{DFM}}|$

Figure 5(a) indicates that $|S_{\psi^-}|$ decreases as $s$ decreases. When the measurement-strength is minimized at $s = 0.21$ [39], we estimate $|S_{\psi^-}|$ as 1.64(4). This is the experimentally observed approximate value of $|S_{\text{DFM}}|$ and is lower than 2, i.e., the classical bound of the CHSH inequality. Because this is an experimental verification of one case (for MES), it gives a loose answer to our question, indicating that a quantum state *can* satisfy the CHSH inequality when the measurement is performed with vanishing disturbance (DFM condition: $s \to 0$ and post-selection to the initial state).

### D. Discussions on Figure 5

Although the $|S|$ of $s = 0.21$ (the weakest measurement performed in our experiment) in Figure 5(a) is our main result, there are several issues worth discussing in Figure 5.

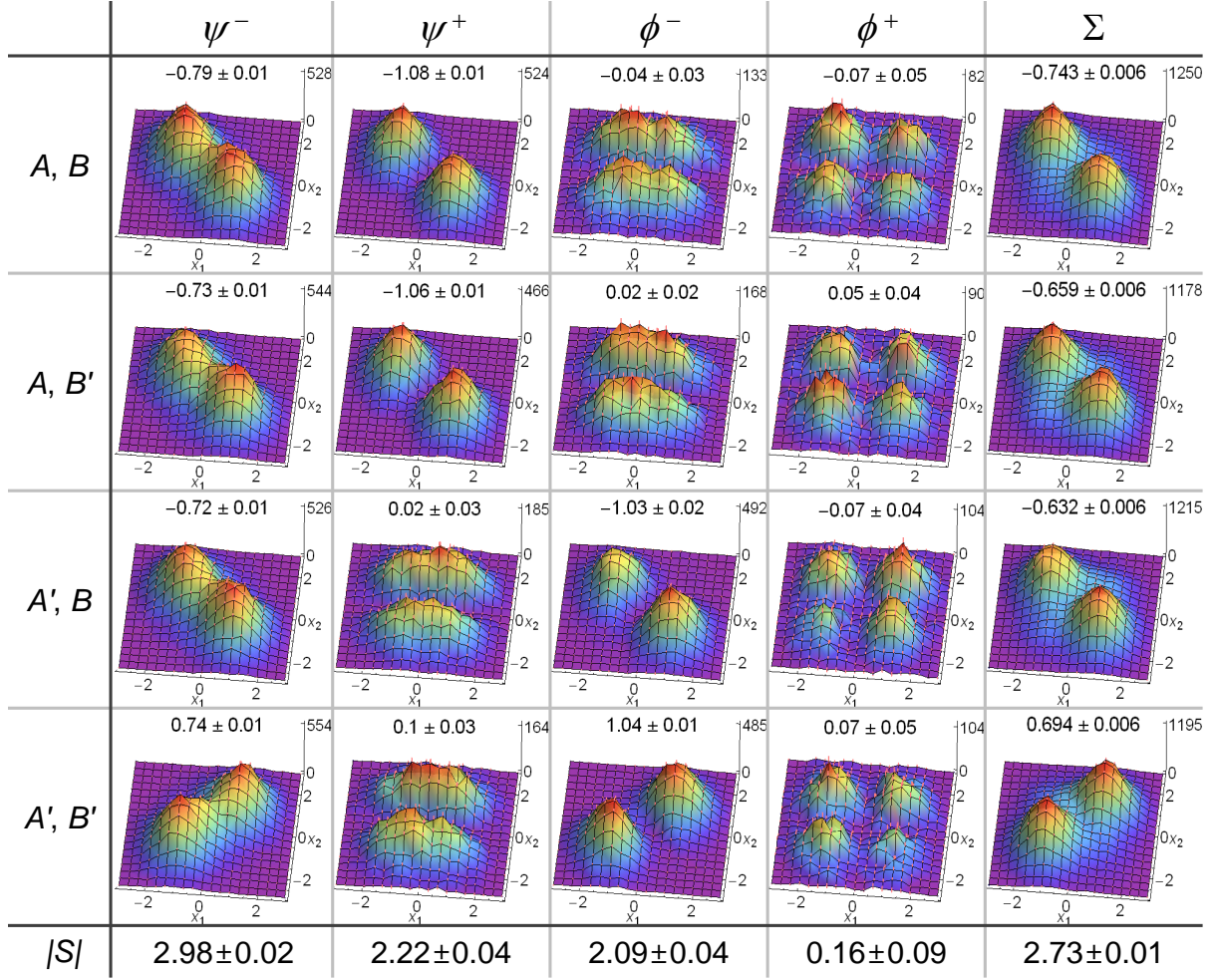

FIG. 4. Probability distributions of pointer states, $\langle AB \rangle$, and $|S|$ for measurement-strength $s \approx 1$. Experimentally measured net coincidence count (per 1 s) distributions for four post-selected states ($\psi^\pm$, $\phi^\pm$) and without post-selection ($\Sigma$). The Poissonian photon-counting errors ($\simeq \sqrt{N}$) are represented by pink error bars. The maximum count in each graph is indicated by the $z$ axis. The expectation values of the $AB$ correlation are given at the top of each graph, with the standard error. The $|S|$ values obtained by adding these values for each column, which correspond to the $|S|$ values of the results post-selected to the state indicated at the top of the column, are presented at the bottom.

1. *Violation of $|S_\Sigma|$ regardless of s in Figure 5(e)*

As $|S_\Sigma|$ represents the result without post-selections, theoretically, it should achieve the maximum violation of the CHSH inequality, $2\sqrt{2}$, for the Bell's state when $s \gg 1$. Figure 5(e) shows that $|S_\Sigma|$ is independent of $s$. Even for the weak limit ($s \ll 1$), the $|S_\Sigma|$ values violate the CHSH inequality by $2\sqrt{2}$, the same amount as the maximum violation of the CHSH inequality. This shows that the quantumness of the initial Bell state is even revealed by weak measurement due to MID, regardless of how small it is. The expectation value of a measurement for an arbitrary $s$ without post-selection is simply equivalent to that of strong measurement (see Appendix B).

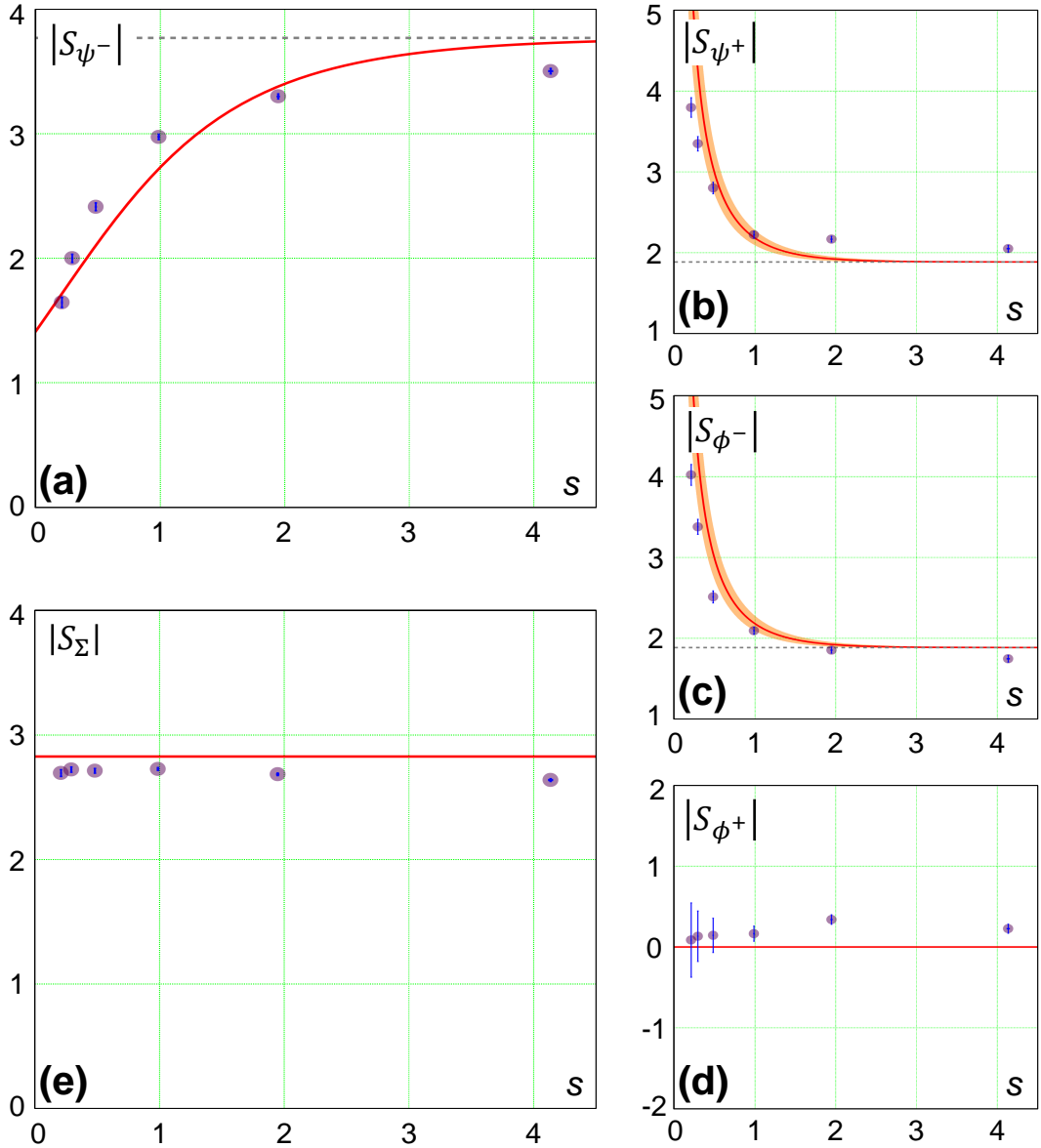

FIG. 5. $|S|$ values post-selected to (a) $\psi^-$, (b) $\psi^+$, (c) $\phi^-$, and (d) $\phi^+$ as functions of $s$, obtained both experimentally (points with error bars) and theoretically (red lines). (e) presents the $|S|$ obtained without post-selection. At the smallest $s$ of 0.21, even though it is not the weak limit ($s \simeq 0$), the $|S|$ of (a) becomes lower than the classical bound, while the $|S|$ of (e) always violates the CHSH inequality.

2. *Post-selected on $|\phi^+\rangle$ in Figure 5(d)*

The detection probability (pointer state) distributions for $|\phi^+\rangle$ in Figure 4 (or 7) have $\pi/2$-rotation symmetry; thus, their expectation values for $\langle x_1 x_2 \rangle$ are zero. Theoretically, the expectation value of a joint-observable for arbitrary $s$ with post-selection to the orthogonal state is expressed by equation (C20). The observables $A(A')$ and $B(B')$ are defined in the $x$-$z$ plane of the Bloch sphere, as described above. On the other hand, $|\phi^+\rangle$ is equivalent to the initial state $|\psi^-\rangle$ under the local unitary operation $\hat{Y}$, i.e., a Pauli operator along the $y$ axis. In the case of $|\psi^-\rangle$ and $|\phi^+\rangle$ with the observables $A(A')$ and $B(B')$, the numerator of equation (C20) is always zero. Therefore, $|S_{\phi^+}| = 0$ regardless of $s$, as shown in Figure 5(d).

3. *Divergences of $|S_{\psi^+}|$ and $|S_{\phi^-}|$ at $s \ll 1$ in Figures 5(b) and (c)*

Similar to $|\phi^+\rangle$, the post-selected states $|\psi^+\rangle$ and $|\phi^-\rangle$ are equivalent to the initial state $|\psi^-\rangle$ under the local unitary operations $\hat{Z}$ and $\hat{X}$, respectively. In these cases, the numerators of equation (C20) are finite, but the denominators, shown in equation (C18), become zero when $s \to 0$. Thus, $|S_{\psi^+}|$ and $|S_{\phi^-}|$ diverge as $s$ vanishes in Figures 5(b) and (c). Considering the definitions of $A(A')$ and $B(B')$, the four pointer-state distributions of $|\psi^+\rangle$ and $|\phi^-\rangle$ in Figure 4 (or 7) are symmetric to each other; thus, the theoretically predicted lines in Figures 5(b) and (c) are the same.

4. *Asymptotic values of $|S_{\psi^-}|$, $|S_{\psi^+}|$, and $|S_{\phi^-}|$ for $s \gg 1$*

In Figure 5, as $s \to \infty$, both the theoretical and experimental $|S_{\psi^-}|$ ($|S_{\psi^+,\phi^-}|$) converge to $8\sqrt{2}/3$ ($4\sqrt{2}/3$), which is even larger than $2\sqrt{2}$, the quantum-mechanical bound of the MES ($< 2$, i.e., the classical bound). This is not surprising considering the physical process for obtaining $|S|$. First, the correlation, such as $\langle AB \rangle$, is acquired via projection measurement ($s \gg 1$). Second, the already strongly measured (disturbed) state is projected (post-selected) to one of the four Bell states, again. The maximum violation of the CHSH inequality is obtained if we calculate $|S|$ from the correlations at the first strong measurement. However, we calculate $|S|$ from the correlations for the case where that the outcome of the post-selection exhibits $\psi^-$, $\psi^+$, and $\phi^-$ at the second measurement. The $|S_{\psi^-}|$, $|S_{\psi^+}|$, and $|S_{\phi^-}|$ obtained in this way (sequential strong measurements) are conditional averages with redundant sampling. These situations differ from DFM. In the DFM condition, the correlation, such as $\langle AB \rangle$, is measured in the weak limit ($s \to 0$); however, in the above cases, the correlations are strongly measured ($s \gg 1$).

### E. Probability of obtaining non-disturbed state in experimental weak limit

We experimentally obtained $p_{\psi^-} = 0.62$, $p_{\psi^+} = 0.18$, $p_{\phi^-} = 0.17$, and $p_{\phi^+} = 0.03$ for the smallest $s$ value of 0.21. This chasm with the theoretical probabilities in the weak limit, $p_{\psi^-} \to 1$ and $p_i \to 0$ ($i = \psi^+, \phi^-, \phi^+$) at $s \to 0$, originates from the following realistic restrictions. First, it is difficult to obtain $s \to 0$ in experiment, because an infinite number of measurements are required in order to obtain a meaningful result in this limit. Thus, $s$ must be finite in real experiments. Second, our post-selection suffers from experimental errors, such as imperfections of the state preparation and Bell-state analyzer. These problems cause degradation of $p_{\psi^-}$ from 1.

### F. Experimental imperfections

Finally, we discuss the experimental errors. First, the strengths of the two weak measurements of $A$ and $B$, which are assumed to be equal, differ within $\delta s = |s_A - s_B| \leq 0.15$, except when $s = 4.14$ ($\delta s \sim 0.33$). This introduces errors in the theoretical curves, as indicated by the orange-shaded regions in Figures 5(b) and (c), but they are negligible in Figures 5(a), (d), and (e). Second, the experimental errors, represented as error bars in the figures, only propagate from the Poissonian error of the photon-counting raw data. In Figures 5(b)–(d), a smaller $s$ yields larger errors. The reason is that the success probability of the post-selection decreases as $s$ decreases. Third, the non-ideal input state is a major factor in the difference between theoretical and experimental values. For an example, in Figure 5(e), the experimental results constantly deviate from the theoretical value. The ratio of the average experimental value to the ideal value ($2\sqrt{2}$) is 95%, which is consistent with the visibility of correlation function of the input state. Fourth, unquantified experimental errors such as wavefront distortion and non-ideal parallelism due to the optics in the weak-measurement apparatus exist; we surmise that these unexpected state-disturbance errors have additional effects on the non-ideal success probabilities for the Bell-state analyzer and the differences between the experimental data and theoretical predictions.

## V. CONCLUSION

We have shown, both theoretically and experimentally, that the outcomes of weak measurement of the initial Bell state post-selected to the initial state meet a necessary condition of classical behavior, in the sense that they satisfy the CHSH inequality. Without post-selection or summing over all possible post-selection outcomes, the results violate the CHSH inequality. This finding implies that, if one measures a quantum system without disturbance, the measurement outcomes can be described by the local hidden variable model. That is, measurement-induced disturbance is a core aspect of quantum mechanics, regardless of how small it is.

## ACKNOWLEDGEMENT

This work was supported by National Research Foundation of Korea (NRF) grants (Nos. 2014R1A1A2054719, 2015R1A2A1A05001819, 2016R1A2B4015978, 2016R1A4A1008978).

**Appendix A: Measurement-induced disturbance**

Here, we describe our weak-measurement procedure. For simplicity, we consider a two-state system and the measurement device described by observables $\hat{A}$ and $\hat{P}$, respectively. The interaction between the two-state system and the device is expressed as

$$\hat{H} = g\delta(t - t_0)\hat{A}\hat{P}, \tag{A1}$$

where $g$ denotes the coupling constant and $\delta(\cdot)$ is a Dirac-delta function implying that the measurement is instantaneously performed at time $t_0$. The device state described by the Gaussian probability distribution

$$|\psi_d(x)|^2 = |\langle x|\psi_d\rangle|^2 = \frac{1}{n^2}e^{-\frac{x^2}{2\sigma^2}}, \tag{A2}$$

where $n = \sqrt[4]{2\pi}\sqrt{\sigma}$, and $\sigma$ is the Gaussian width, moves along the $x$ axis depending on the $\hat{A}$ value of the system, as $\hat{P}$ is the momentum operator of the measurement device. $x$ acts as the device pointer. If the system initially lies at $|\psi_s\rangle = \alpha|1\rangle_A + \beta|-1\rangle_A$, where $|1\rangle_A$ and $|-1\rangle_A$ denote the bases of $\hat{A}$, the total system after interaction with the device is expressed as

$$\langle x|\Psi\rangle = \alpha|1\rangle\frac{1}{n}e^{-\frac{(x-g)^2}{4\sigma^2}} + \beta|-1\rangle\frac{1}{n}e^{-\frac{(x+g)^2}{4\sigma^2}}. \tag{A3}$$

By tracing the device, we obtain the system density matrix

$$\rho'_s = \text{Tr}_d|\Psi\rangle\langle\Psi| = \begin{pmatrix} |\alpha|^2 & \alpha\beta^*\gamma \\ \alpha^*\beta\gamma & |\beta|^2 \end{pmatrix}, \tag{A4}$$

where $\gamma$ represents the overlap between two device states moving to $+g$ and $-g$, that is,

$$\gamma = \int \frac{1}{n^2}e^{-\frac{(x-g)^2}{4\sigma^2}}e^{-\frac{(x+g)^2}{4\sigma^2}}dx = e^{-\frac{g^2}{2\sigma^2}} = e^{-s}. \tag{A5}$$

Here, the measurement strength $s$ is introduced [23]. Below, we set $g = 1$, as $s$ is determined from the ratio of $g$ to $\sigma$.

$\rho'_s$ is decomposed into

$$\rho'_s = \frac{1+\gamma}{2}|\psi_s\rangle\langle\psi_s| + \frac{1-\gamma}{2}|\bar{\psi}_s\rangle\langle\bar{\psi}_s|, \tag{A6}$$

where $|\bar{\psi}_s\rangle = \alpha|1\rangle_A - \beta|-1\rangle_A$, i.e., a mirror-symmetric state with $\langle\psi_s|\bar{\psi}_s\rangle = \cos\theta$, as shown in Figure 6. Geometrically, $\rho_s$ is located at a point along a straight line connecting $|\psi_s\rangle$ and $|\bar{\psi}_s\rangle$ in the Bloch sphere. The initial pure-state vector decreases in length in response to the measurement-induced decoherence. After weak measurement, there exists a probability that the state continues to survive in the initial state, which is expressed as $(1+\gamma)/2$. We then find the disturbance

$$D = 1 - \langle\psi_s|\rho'_s|\psi_s\rangle = \frac{1-\gamma}{2}\sin^2\theta \leq \frac{1-\gamma}{2}. \tag{A7}$$

At the weak-measurement limit, i.e., $s \ll 1$, we obtain $D < s$, implying that the disturbance is negligible.

The above argument is also applied to higher-dimensional states. After measurement, the state is decomposed into

$$|\psi_s\rangle\langle\psi_s| \to (1-p)|\psi_s\rangle\langle\psi_s| + p\sum_k c_k|\bar{\psi}_k\rangle\langle\bar{\psi}_k|, \tag{A8}$$

where $\sum_k c_k = 1$. $p$ must be minimized, as such decomposition is not unique. We then obtain

$$D \leq \min\{p\}. \tag{A9}$$

In the weak-measurement limit, we intuitively recognize that $p \to 0$.

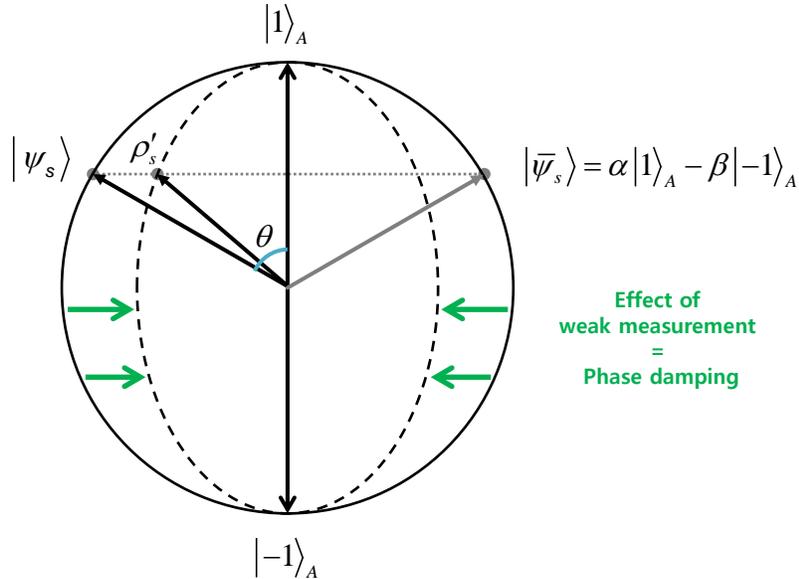

FIG. 6. Depiction of system qubit state during measurement.

**Appendix B: Expectation value of observable**

Here, we show that the average of the weak-measurement outcomes described in Appendix A are equivalent to the quantum-mechanical average. After the interaction between the system and the device, the total system is described by equation (A3). By tracing the system, we obtain the device-pointer probability distribution:

$$p(x) = \frac{1}{n^2}\left(|\alpha|^2 e^{-\frac{(x-1)^2}{2\sigma^2}} + |\beta|^2 e^{-\frac{(x+1)^2}{2\sigma^2}}\right). \tag{B1}$$

The average of the pointer positions is equal to the expectation value of the $\hat{A}$ operator, such that

$$\langle x \rangle = \int x \, p(x) dx = |\alpha|^2 - |\beta|^2 = \langle \hat{A} \rangle. \tag{B2}$$

We emphasize that equation (B2) does not depend on the measurement uncertainty $\sigma$ (or $s$), because the center of the pointer probability distribution remains intact regardless of the magnitude of $\sigma$. In conventional experiments, this is acceptable, because the center values of moved distributions correspond to the eigenvalues. We will show that equation (B2) is also true in our experiment. This argument has been applied to more generalized cases, such as multidimensional and multi-partite systems and weak values incorporating post-selection [23].

**Appendix C: Expectation value of joint observables**

Here, we show the calculation method for the expectation value of joint observables $\hat{A}$ and $\hat{B}$. Our strategy is as follows. First, we find the device density matrix $\rho'_d$, measuring $\hat{A}$ and $\hat{B}$. The corresponding pointers of these observables are $\hat{X}_1$ and $\hat{X}_2$, respectively, when both weak measurement and post-selection are completed. Then, we calculate the average of $\hat{X}_1$ and $\hat{X}_2$, i.e., $\text{Tr}(\hat{X}_1\hat{X}_2\rho'_d)$.

The interaction Hamiltonian is expressed as

$$\hat{H} = g_1\delta(t - t_1)\hat{A}\hat{P}_1 + g_2\delta(t - t_2)\hat{B}\hat{P}_2, \tag{C1}$$

where $g_1$ and $g_2$ are coupling constants; $\hat{P}_1$ and $\hat{P}_2$ are the momentum operators related to the pointer-state position operators $\hat{X}_1$ and $\hat{X}_2$, respectively; and $t_i$ is the time until the measurement. Equation (C1) indicates that the two measurements for $A$ and $B$ are independent, which can be applied to spatially or temporally separated cases. We assume that $g_1 = g_2 = g$ for simplicity and that $\hat{A}$ and $\hat{B}$ satisfy $\hat{A}^2 = \hat{B}^2 = I$, where $I$ is the identity. The initial system state is an arbitrary two-qubit state denoted as $|\psi_i\rangle$, and the measurement device is prepared as a wavefunction of the normalized two-dimensional Gaussian distribution

$$\langle x_1, x_2 | \Phi \rangle = \Phi(x_1, x_2) = \frac{1}{\sqrt{2\pi\sigma_1\sigma_2}} \exp\left(-\frac{x_1^2}{4\sigma_1^2}\right) \exp\left(-\frac{x_2^2}{4\sigma_2^2}\right), \tag{C2}$$

where $\sigma_1$ and $\sigma_2$ are the probability distribution widths in $x_1$ and $x_2$, respectively. After the interaction between the system and the measurement device, which is described by the unitary evolution $U = e^{-ig(\hat{A}\hat{P}_1 + \hat{B}\hat{P}_2)}$, the total system state is expressed as

$$\begin{aligned}
\rho' &\equiv U\rho U^\dagger, \\
&= \rho + \sum_{n=1}^{\infty} \frac{(-ig)^n}{n!} \mathbf{ad}^n[\hat{A}\hat{P}_1 \circ \rho] + \sum_{m=1}^{\infty} \frac{(-ig)^m}{m!} \mathbf{ad}^m[\hat{B}\hat{P}_2 \circ \rho] \\
&+ \sum_{n,m=1}^{\infty} \frac{(-ig)^n}{n!} \frac{(-ig)^m}{m!} \mathbf{ad}^n[\hat{A}\hat{P}_1 \circ (\mathbf{ad}^m[\hat{B}\hat{P}_2 \circ \rho])],
\end{aligned} \tag{C3}$$

where $\rho$ is the initial total density matrix, $\mathbf{ad}^1[\Omega \circ \Theta] \equiv [\Omega, \Theta]$, and $\mathbf{ad}^n[\Omega \circ \Theta] \equiv \mathbf{ad}[\Omega \circ (\mathbf{ad}^{n-1}[\Omega \circ \Theta])] = [\Omega, \mathbf{ad}^{n-1}[\Omega \circ \Theta]]$. By tracing the measurement device, we obtain the system density matrix

$$\begin{aligned}
\rho'_s &= Tr_d\rho', \\
&= \rho_s + \sum_{n=1}^{\infty} \frac{(-ig)^n \langle \hat{P}_1^n \rangle}{n!} \mathbf{ad}^n[\hat{A} \circ \rho_s] + \sum_{m=1}^{\infty} \frac{(-ig)^m \langle \hat{P}_2^m \rangle}{m!} \mathbf{ad}^m[\hat{B} \circ \rho_s] \\
&+ \sum_{n,m=1}^{\infty} \frac{(-ig)^n \langle \hat{P}_1^n \rangle}{n!} \frac{(-ig)^m \langle \hat{P}_2^m \rangle}{m!} \mathbf{ad}^n[\hat{A} \circ (\mathbf{ad}^m[\hat{B} \circ \rho_s])],
\end{aligned} \tag{C4}$$

where $\langle \hat{P}^n \rangle \equiv Tr_d(\hat{P}^n \rho_d)$.

By tracing the system, the measurement-device density matrix after post-selection on the state $|\psi_f\rangle$ is expressed as

$$\rho'_d = \frac{Tr_s \rho' \Pi_f}{Tr \rho' \Pi_f}, \tag{C5}$$

where $\Pi_f = |\psi_f\rangle\langle\psi_f|$. To calculate equation (C5) in the form of the weak value, we rewrite this expression as

$$\rho'_d = \frac{Z\rho'_d}{Z}, \tag{C6}$$

where $Z = Tr\rho'\Pi_f/|\langle\psi_f|\psi_i\rangle|^2$ and $Z\rho'_d = Tr_s\rho'\Pi_f/|\langle\psi_f|\psi_i\rangle|^2$, if $\langle\psi_f|\psi_i\rangle \neq 0$. Below, we calculate $Z$ and $Z\rho'_d$ separately. The case of $\langle\psi_f|\psi_i\rangle = 0$ is considered in subsection C 2.

### 1. Non-orthogonal case

From equation (C3), we obtain

$$Z = 1 + \sum_{n=1}^{\infty} \frac{(-ig)^n \langle \hat{P}_1^n \rangle}{n!} \sum_{j=0}^{n} (-1)^j {}_nC_j \langle \hat{A}^{n-j} \rangle_w \langle \hat{A}^j \rangle_w^* + \sum_{m=1}^{\infty} \frac{(-ig)^m \langle \hat{P}_2^m \rangle}{m!} \sum_{k=0}^{m} (-1)^k {}_mC_k \langle \hat{B}^{m-k} \rangle_w \langle \hat{B}^k \rangle_w^*$$

$$+ \sum_{n,m=1}^{\infty} \frac{(-ig)^n \langle \hat{P}_1^n \rangle}{n!} \frac{(-ig)^m \langle \hat{P}_2^m \rangle}{m!} \sum_{j=0}^{n} (-1)^j {}_nC_j \sum_{k=0}^{m} (-1)^k {}_mC_k \langle \hat{A}^{n-j} \hat{B}^{m-k} \rangle_w \langle \hat{A}^j \hat{B}^k \rangle_w^*, \tag{C7}$$

$$= Z_1 + Z_2, \tag{C8}$$

where $s_i \equiv g^2/(2\sigma_i^2) = 2g^2 \langle \hat{P}_i^2 \rangle$ is the strength of the weak measurement, $\langle O \rangle_w = \langle \psi_f | O | \psi_i \rangle / \langle \psi_f | \psi_i \rangle$,

$$Z_1 = 1 + \frac{1}{2}(1 - |\langle \hat{A} \rangle_w|^2)(e^{-s_1} - 1) + \frac{1}{2}(1 - |\langle \hat{B} \rangle_w|^2)(e^{-s_2} - 1), \tag{C9}$$

and

$$Z_2 = \Big[ \Big( \sum_{n=1}^{\infty} \frac{(-ig)^{2n} \langle \hat{P}_1^{2n} \rangle}{(2n)!} \sum_{j=0}^{2n} (-1)^j {}_{2n}C_j \Big) \Big( \sum_{m=1}^{\infty} \frac{(-ig)^{2m} \langle \hat{P}_2^{2m} \rangle}{(2m)!} \sum_{k=0}^{2m} (-1)^k {}_{2m}C_k \Big) \langle \hat{A}^{2n-j} \hat{B}^{2m-k} \rangle_w \langle \hat{A}^j \hat{B}^k \rangle_w^*$$

$$+ \Big( \sum_{n=1}^{\infty} \frac{(-ig)^{2n} \langle \hat{P}_1^{2n} \rangle}{(2n)!} \sum_{j=0}^{2n} (-1)^j {}_{2n}C_j \Big) \Big( \sum_{m=0}^{\infty} \frac{(-ig)^{2m+1} \langle \hat{P}_2^{2m+1} \rangle}{(2m+1)!} \sum_{k=0}^{2m+1} (-1)^k {}_{2m+1}C_k \Big) \langle \hat{A}^{2n-j} \hat{B}^{2m+1-k} \rangle_w \langle \hat{A}^j \hat{B}^k \rangle_w^*$$

$$+ \Big( \sum_{n=0}^{\infty} \frac{(-ig)^{2n+1} \langle \hat{P}_1^{2n+1} \rangle}{(2n+1)!} \sum_{j=0}^{2n+1} (-1)^j {}_{2n+1}C_j \Big) \Big( \sum_{m=1}^{\infty} \frac{(-ig)^{2m} \langle \hat{P}_2^{2m} \rangle}{(2m)!} \sum_{k=0}^{2m} (-1)^k {}_{2m}C_k \Big) \langle \hat{A}^{2n+1-j} \hat{B}^{2m-k} \rangle_w \langle \hat{A}^j \hat{B}^k \rangle_w^*$$

$$+ \Big( \sum_{n=0}^{\infty} \frac{(-ig)^{2n+1} \langle \hat{P}_1^{2n+1} \rangle}{(2n+1)!} \sum_{j=0}^{2n+1} (-1)^j {}_{2n+1}C_j \Big) \Big( \sum_{m=0}^{\infty} \frac{(-ig)^{2m+1} \langle \hat{P}_2^{2m+1} \rangle}{(2m+1)!} \sum_{k=0}^{2m+1} (-1)^k {}_{2m+1}C_k \Big)$$

$$\times \langle \hat{A}^{2n+1-j} \hat{B}^{2m+1-k} \rangle_w \langle \hat{A}^j \hat{B}^k \rangle_w^* \Big]. \tag{C10}$$

Using $\langle \hat{P}_i^{2n+1} \rangle = 0$, $\langle \hat{P}_i^{2n} \rangle = (2n-1)!! \langle \hat{P}_i^2 \rangle^n$ and $\sum_{j=0}^{n} {}_{2n}C_{2j} = \sum_{j=0}^{n-1} {}_{2n}C_{2j+1} = 2^{2n-1}$, $Z_2$ is rewritten as

$$Z_2 = \sum_{n=1}^{\infty} \frac{(-ig)^{2n}(2n-1)!! \langle \hat{P}_1^2 \rangle^n}{(2n)!} \sum_{j=0}^{2n} (-1)^j {}_{2n}C_j \sum_{m=1}^{\infty} \frac{(-ig)^{2m}(2m-1)!! \langle \hat{P}_2^2 \rangle^m}{(2m)!} \sum_{k=0}^{2m} (-1)^k {}_{2m}C_k \langle \hat{A}^{2n-j} \hat{B}^{2m-k} \rangle_w \langle \hat{A}^j \hat{B}^k \rangle_w^*,$$

$$= \sum_{n=1}^{\infty} \frac{(-1)^n (g)^{2n} \langle \hat{P}_1^2 \rangle^n}{2^n n!} \sum_{m=1}^{\infty} \frac{(-1)^m (g)^{2m} \langle \hat{P}_2^2 \rangle^m}{2^m m!} (2^{2n+2m-2})(1 - |\langle \hat{A} \rangle_w|^2 - |\langle \hat{B} \rangle_w|^2 + |\langle \hat{A}\hat{B} \rangle_w|^2),$$

$$= \frac{1}{4}(1 - |\langle \hat{A} \rangle_w|^2 - |\langle \hat{B} \rangle_w|^2 + |\langle \hat{A}\hat{B} \rangle_w|^2)(e^{-s_1} - 1)(e^{-s_2} - 1). \tag{C11}$$

We finally obtain

$$Z = 1 + \frac{1}{2}(1 - |\langle \hat{A} \rangle_w|^2)(e^{-s_1} - 1) + \frac{1}{2}(1 - |\langle \hat{B} \rangle_w|^2)(e^{-s_2} - 1)$$

$$+ \frac{1}{4}\big(1 - |\langle \hat{A} \rangle_w|^2 - |\langle \hat{B} \rangle_w|^2 + |\langle \hat{A}\hat{B} \rangle_w|^2 \big)\big(e^{-s_1} - 1\big)\big(e^{-s_2} - 1\big). \tag{C12}$$

Note that $Z \to 1$ as $s_{1,2} \to 0$.

Now, we calculate $Z\rho_d'$. Using equations (C3), (C5), and (C6), we obtain

$$
\begin{aligned}
Z\rho_d' = \Big[ &\rho_d^1 \rho_d^2 + \sum_{n,m=1}^{\infty} \frac{(-ig)^{2n}(-ig)^{2m}}{(2n)!(2m)!} \Big( \sum_{j=0}^{n} {}_{2n}C_{2j} \hat{P}_1^{2n-2j} \rho_d^1 \hat{P}_1^{2j} \sum_{k=0}^{m} {}_{2m}C_{2k} \hat{P}_2^{2m-2k} \rho_d^2 \hat{P}_2^{2k} \\
&- |\langle \hat{A} \rangle_w|^2 \sum_{j=0}^{n} {}_{2n}C_{2j} \hat{P}_1^{2n-2j} \rho_d^1 \hat{P}_1^{2j} \sum_{k=0}^{m-1} {}_{2m}C_{2k+1} \hat{P}_2^{2m-2k-1} \rho_d^2 \hat{P}_2^{2k+1} \\
&- |\langle \hat{B} \rangle_w|^2 \sum_{j=0}^{n-1} {}_{2n}C_{2j+1} \hat{P}_1^{2n-2j-1} \rho_d^1 \hat{P}_1^{2j+1} \sum_{k=0}^{m} {}_{2m}C_{2k} \hat{P}_2^{2m-2k} \rho_d^2 \hat{P}_2^{2k} \\
&+ |\langle \hat{A}\hat{B} \rangle_w|^2 \sum_{j=0}^{n-1} {}_{2n}C_{2j+1} \hat{P}_1^{2n-2j-1} \rho_d^1 \hat{P}_1^{2j+1} \sum_{k=0}^{m-1} {}_{2m}C_{2k+1} \hat{P}_2^{2m-2k-1} \rho_d^2 \hat{P}_2^{2k+1} \Big) \\
&+ \sum_{n=1}^{\infty} \frac{(-ig)^{2n}}{(2n)!} \sum_{m=0}^{\infty} \frac{(-ig)^{2m+1}}{(2m+1)!} \Big( \langle \hat{B} \rangle_w \sum_{j=0}^{n} {}_{2n}C_{2j} \hat{P}_1^{2n-2j} \rho_d^1 \hat{P}_1^{2j} \sum_{k=0}^{m} {}_{2m+1}C_{2k} \hat{P}_2^{2m+1-2k} \rho_d^2 \hat{P}_2^{2k} \\
&- \langle \hat{B} \rangle_w^* \sum_{j=0}^{n} {}_{2n}C_{2j} \hat{P}_1^{2n-2j} \rho_d^1 \hat{P}_1^{2j} \sum_{k=0}^{m-1} {}_{2m+1}C_{2k+1} \hat{P}_2^{2m-2k} \rho_d^2 \hat{P}_2^{2k+1} \\
&- \langle \hat{A}\hat{B} \rangle_w \langle \hat{A} \rangle_w^* \sum_{j=0}^{n-1} {}_{2n}C_{2j+1} \hat{P}_1^{2n-2j-1} \rho_d^1 \hat{P}_1^{2j+1} \sum_{k=0}^{m} {}_{2m+1}C_{2k} \hat{P}_2^{2m+1-2k} \rho_d^2 \hat{P}_2^{2k} \\
&+ \langle \hat{A} \rangle_w \langle \hat{A}\hat{B} \rangle_w^* \sum_{j=0}^{n-1} {}_{2n}C_{2j+1} \hat{P}_1^{2n-2j-1} \rho_d^1 \hat{P}_1^{2j+1} \sum_{k=0}^{m-1} {}_{2m+1}C_{2k+1} \hat{P}_2^{2m-2k} \rho_d^2 \hat{P}_2^{2k+1} \Big) \\
&+ \sum_{n=0}^{\infty} \frac{(-ig)^{2n+1}}{(2n+1)!} \sum_{m=1}^{\infty} \frac{(-ig)^{2m}}{(2m)!} \Big( \langle \hat{A} \rangle_w \sum_{j=0}^{n} {}_{2n+1}C_{2j} \hat{P}_1^{2n+1-2j} \rho_d^1 \hat{P}_1^{2j} \sum_{k=0}^{m} {}_{2m}C_{2k} \hat{P}_2^{2m-2k} \rho_d^2 \hat{P}_2^{2k} \\
&- \langle \hat{A}\hat{B} \rangle_w \langle \hat{B} \rangle_w^* \sum_{j=0}^{n} {}_{2n+1}C_{2j} \hat{P}_1^{2n+1-2j} \rho_d^1 \hat{P}_1^{2j} \sum_{k=0}^{m-1} {}_{2m}C_{2k+1} \hat{P}_2^{2m-2k-1} \rho_d^2 \hat{P}_2^{2k+1} \\
&- \langle \hat{A} \rangle_w^* \sum_{j=0}^{n-1} {}_{2n+1}C_{2j+1} \hat{P}_1^{2n-2j} \rho_d^1 \hat{P}_1^{2j+1} \sum_{k=0}^{m} {}_{2m}C_{2k} \hat{P}_2^{2m-2k} \rho_d^2 \hat{P}_2^{2k} \\
&+ \langle \hat{B} \rangle_w \langle \hat{A}\hat{B} \rangle_w^* \sum_{j=0}^{n-1} {}_{2n+1}C_{2j+1} \hat{P}_1^{2n-2j} \rho_d^1 \hat{P}_1^{2j+1} \sum_{k=0}^{m-1} {}_{2m}C_{2k+1} \hat{P}_2^{2m-2k-1} \rho_d^2 \hat{P}_2^{2k+1} \Big) \\
&+ \sum_{n=0}^{\infty} \frac{(-ig)^{2n+1}}{(2n+1)!} \sum_{m=0}^{\infty} \frac{(-ig)^{2m+1}}{(2m+1)!} \Big( \langle \hat{A}\hat{B} \rangle_w \sum_{j=0}^{n} {}_{2n+1}C_{2j} \hat{P}_1^{2n+1-2j} \rho_d^1 \hat{P}_1^{2j} \sum_{k=0}^{m} {}_{2m+1}C_{2k} \hat{P}_2^{2m+1-2k} \rho_d^2 \hat{P}_2^{2k} \\
&- \langle \hat{A} \rangle_w \langle \hat{B} \rangle_w^* \sum_{j=0}^{n} {}_{2n+1}C_{2j} \hat{P}_1^{2n+1-2j} \rho_d^1 \hat{P}_1^{2j} \sum_{k=0}^{m-1} {}_{2m+1}C_{2k+1} \hat{P}_2^{2m-2k} \rho_d^2 \hat{P}_2^{2k+1} \\
&- \langle \hat{B} \rangle_w \langle \hat{A} \rangle_w^* \sum_{j=0}^{n-1} {}_{2n+1}C_{2j+1} \hat{P}_1^{2n-2j} \rho_d^1 \hat{P}_1^{2j+1} \sum_{k=0}^{m} {}_{2m+1}C_{2k} \hat{P}_2^{2m+1-2k} \rho_d^2 \hat{P}_2^{2k} \\
&+ \langle \hat{A}\hat{B} \rangle_w^* \sum_{j=0}^{n-1} {}_{2n+1}C_{2j+1} \hat{P}_1^{2n-2j} \rho_d^1 \hat{P}_1^{2j+1} \sum_{k=0}^{m-1} {}_{2m+1}C_{2k+1} \hat{P}_2^{2m-2k} \rho_d^2 \hat{P}_2^{2k+1} \Big) \Big].
\end{aligned}
\tag{C13}
$$

The pointer-position expectation values in the weak-value scheme discussed above are expressed as

$$\langle \hat{X}_j \rangle_{\psi_f, \psi_i} \equiv Tr(\hat{X}_j \rho_d'), \qquad \langle \hat{X}_1 \hat{X}_2 \rangle_{\psi_f, \psi_i} \equiv Tr(\hat{X}_1 \hat{X}_2 \rho_d'). \tag{C14}$$

Using equations (C13) and (C14), $\langle \hat{X}\hat{P}^n + \hat{P}^n \hat{X} \rangle = 0$ (for $n \geq 1$), and $[\hat{X}, \hat{P}] = i$, we find

$$\langle \hat{X}_i \hat{X}_j \rangle_{\psi_f \not\perp \psi_i} = \frac{1}{2Z} (\mathrm{Re}[\langle \hat{A}\hat{B} \rangle_w + \langle \hat{A} \rangle_w \langle \hat{B} \rangle_w^*]), \tag{C15}$$

where we set $g = 1$ without loss of generality.

To achieve measurement without disturbing the system, we simply set the final postselected state to the initial state, as the undisturbed state should correspond to the original state. Thus, the outcome of the DFM of two observables is defined as

$$\langle \hat{A}\hat{B} \rangle_{\text{DFM}} \equiv \langle \hat{X}_i \hat{X}_j \rangle_{\psi_i, \psi_i} = \frac{1}{2}\left( \langle \hat{A}\hat{B} \rangle + \langle \hat{A} \rangle \langle \hat{B} \rangle \right). \tag{C16}$$

Note that in equation (C16), the weak value $\langle \cdot \rangle_w$ is replaced by the average $\langle \cdot \rangle$.

## 2. Orthogonal case

If $\langle \psi_f | \psi_i \rangle = 0$, we must calculate $Z$ and $Z\rho'_d$ in a slightly different way. The denominator of equation (C5) is calculated as

$$\begin{aligned}
Tr\rho'\Pi_f &= \sum_{n=1}^{\infty} \frac{(-ig)^n \langle \hat{P}_1^n \rangle}{n!} \sum_{j=0}^{n} (-1)^j {}_nC_j \langle \psi_f | \hat{A}^{n-j} | \psi_i \rangle \langle \psi_i | \hat{A}^j | \psi_f \rangle \\
&+ \sum_{m=1}^{\infty} \frac{(-ig)^m \langle \hat{P}_2^m \rangle}{m!} \sum_{k=0}^{m} (-1)^k {}_mC_k \langle \psi_f | \hat{B}^{m-k} | \psi_i \rangle \langle \psi_i | \hat{B}^k | \psi_f \rangle \\
&+ \sum_{n,m=1}^{\infty} \frac{(-ig)^n \langle \hat{P}_1^n \rangle}{n!} \frac{(-ig)^m \langle \hat{P}_2^m \rangle}{m!} \sum_{j=0}^{n}(-1)^j {}_nC_j \sum_{k=0}^{m}(-1)^k {}_mC_k \langle \psi_f | \hat{A}^{n-j}\hat{B}^{m-k} | \psi_i \rangle \langle \psi_i | \hat{A}^j \hat{B}^k | \psi_f \rangle, \\
&= g^2 \langle \hat{P}_1^2 \rangle |\langle \psi_f | \hat{A} | \psi_i \rangle|^2 \Big[ 1 + \sum_{n=1}^{\infty} \frac{(-ig)^n \langle \hat{P}_1^{n+2} \rangle}{n! \langle \hat{P}_1^2 \rangle} \sum_{j=0}^{n}(-1)^j {}_nC_j \langle \hat{A}^{n-j} \rangle_{ow} \langle \hat{A}^j \rangle_{ow}^* \Big] \\
&+ g^2 \langle \hat{P}_2^2 \rangle |\langle \psi_f | \hat{B} | \psi_i \rangle|^2 \Big[ 1 + \sum_{m=1}^{\infty} \frac{(-ig)^m \langle \hat{P}_2^{m+2} \rangle}{m! \langle \hat{P}_2^2 \rangle} \sum_{k=0}^{m}(-1)^k {}_mC_k \langle \hat{B}^{m-k} \rangle_{ow} \langle \hat{m}^k \rangle_{ow}^* \Big] \\
&+ \sum_{n=1}^{\infty} \frac{(-ig)^{2n} \langle \hat{P}_1^{2n} \rangle}{(2n)!} \sum_{j=0}^{2n}(-1)^j {}_{2n}C_j \sum_{m=1}^{\infty} \frac{(-ig)^{2m} \langle \hat{P}_2^{2m} \rangle}{(2m!)} \sum_{k=0}^{2m}(-1)^k {}_{2m}C_k \langle \psi_f | \hat{A}^{2n-j}\hat{B}^{2m-k} | \psi_i \rangle \langle \psi_i | \hat{A}^j \hat{B}^k | \psi_f \rangle,
\end{aligned} \tag{C17}$$

where $\langle \hat{A}^n \rangle_{ow} \equiv \langle \psi_f | \hat{A}^{n+1} | \psi_i \rangle / \langle \psi_f | \hat{A}(n+1) | \psi_i \rangle$. Using $\langle \hat{P}_i^{2n+1} \rangle = 0$ and $\langle \hat{P}_i^{2n+2} \rangle = (2n+1)!! \langle \hat{P}_i^2 \rangle^{n+1}$, from equation (C17), we obtain

$$\begin{aligned}
Tr\rho'\Pi_f &= \frac{s_1}{2}\Big(1 + \sum_{n=1}^{\infty} \frac{(-ig)^{2n}\langle \hat{P}_1^2 \rangle^n}{(2n+2)!!} \sum_{j=0}^{n} \frac{(2n+2)!}{(2n-2j+1)!(2j+1)!} |\langle \psi_f | \hat{A} | \psi_i \rangle|^2 \Big) \\
&+ \frac{s_2}{2}\Big(1 + \sum_{m=1}^{\infty} \frac{(-ig)^{2m}\langle \hat{P}_2^2 \rangle^m}{(2m+2)!!} \sum_{k=0}^{m} \frac{(2m+2)!}{(2m-2k+1)!(2k+1)!} |\langle \psi_f | \hat{B} | \psi_i \rangle|^2 \Big) \\
&+ \sum_{n,m=1}^{\infty} \frac{(-ig)^{2n}(2n-1)!!\langle \hat{P}_1^2 \rangle^n}{(2n)!} \frac{(-ig)^{2m}(2m-1)!!\langle \hat{P}_2^2 \rangle^m}{(2m)!} \Big( -\sum_{j=0}^{n-1} {}_{2n}C_{2j+1} \sum_{k=0}^{m} {}_{2m}C_{2k} |\langle \psi_f | \hat{A} | \psi_i \rangle|^2 \\
&- \sum_{j=0}^{n} {}_{2n}C_{2j} \sum_{k=0}^{m-1} {}_{2m}C_{2k+1} |\langle \psi_f | \hat{B} | \psi_i \rangle|^2 + \sum_{j=0}^{n-1} {}_{2n}C_{2j+1} \sum_{k=0}^{m-1} {}_{2m}C_{2k+1} |\langle \psi_f | \hat{A}\hat{B} | \psi_i \rangle|^2 \Big), \\
&= \frac{1}{2}|\langle \psi_f | \hat{A} | \psi_i \rangle|^2 (1 - e^{-s_1}) + \frac{1}{2}|\langle \psi_f | \hat{B} | \psi_i \rangle|^2 (1 - e^{-s_2}) \\
&+ \frac{1}{4}(|\langle \psi_f | \hat{A}\hat{B} | \psi_i \rangle|^2 - |\langle \psi_f | \hat{A} | \psi_i \rangle|^2 - |\langle \psi_f | \hat{B} | \psi_i \rangle|^2)(1 - e^{-s_1})(1 - e^{-s_2})
\end{aligned} \tag{C18}$$

The numerator of equation (C5) is calculated to be

$$Tr_s\rho'\Pi_f = \Big[ -\langle\psi_f|\hat{A}|\psi_i\rangle\langle\psi_i|\hat{A}|\psi_f\rangle \sum_{j=0}^{n} {}_{2n}C_{2j} \hat{P}_1^{2n-2j} \rho_d^1 \hat{P}_1^{2j} \sum_{k=0}^{m-1} {}_{2m}C_{2k+1} \hat{P}_2^{2m-2k-1} \rho_d^2 \hat{P}_2^{2k+1}$$

$$- \langle\psi_f|\hat{B}|\psi_i\rangle\langle\psi_i|\hat{B}|\psi_f\rangle \sum_{j=0}^{n-1} {}_{2n}C_{2j+1} \hat{P}_1^{2n-2j-1} \rho_d^1 \hat{P}_1^{2j+1} \sum_{k=0}^{m} {}_{2m}C_{2k} \hat{P}_2^{2m-2k} \rho_d^2 \hat{P}_2^{2k}$$

$$+ \langle\psi_f|\hat{A}\hat{B}|\psi_i\rangle\langle\psi_i|\hat{A}\hat{B}|\psi_f\rangle \sum_{j=0}^{n-1} {}_{2n}C_{2j+1} \hat{P}_1^{2n-2j-1} \rho_d^1 \hat{P}_1^{2j+1} \sum_{k=0}^{m-1} {}_{2m}C_{2k+1} \hat{P}_2^{2m-2k-1} \rho_d^2 \hat{P}_2^{2k+1} \Big)$$

$$+ \sum_{n=1}^{\infty} \frac{(-ig)^{2n}}{(2n)!} \sum_{m=0}^{\infty} \frac{(-ig)^{2m+1}}{(2m+1)!} \Big( -\langle\psi_f|\hat{A}\hat{B}|\psi_i\rangle\langle\psi_i|\hat{A}|\psi_f\rangle \sum_{j=0}^{n-1} {}_{2n}C_{2j+1} \hat{P}_1^{2n-2j-1} \rho_d^1 \hat{P}_1^{2j+1} \sum_{k=0}^{m} {}_{2m+1}C_{2k} \hat{P}_2^{2m+1-2k} \rho_d^2 \hat{P}_2^{2k}$$

$$+ \langle\psi_f|\hat{A}|\psi_i\rangle\langle\psi_i|\hat{A}\hat{B}|\psi_f\rangle \sum_{j=0}^{n-1} {}_{2n}C_{2j+1} \hat{P}_1^{2n-2j-1} \rho_d^1 \hat{P}_1^{2j+1} \sum_{k=0}^{m-1} {}_{2m+1}C_{2k+1} \hat{P}_2^{2m-2k} \rho_d^2 \hat{P}_2^{2k+1} \Big)$$

$$+ \sum_{n=0}^{\infty} \frac{(-ig)^{2n+1}}{(2n+1)!} \sum_{m=1}^{\infty} \frac{(-ig)^{2m}}{(2m)!} \Big( -\langle\psi_f|\hat{A}\hat{B}|\psi_i\rangle\langle\psi_i|\hat{B}|\psi_f\rangle \sum_{j=0}^{n} {}_{2n+1}C_{2j} \hat{P}_1^{2n+1-2j} \rho_d^1 \hat{P}_1^{2j} \sum_{k=0}^{m-1} {}_{2m}C_{2k+1} \hat{P}_2^{2m-2k-1} \rho_d^2 \hat{P}_2^{2k+1}$$

$$+ \langle\psi_f|\hat{B}|\psi_i\rangle\langle\psi_i|\hat{A}\hat{B}|\psi_f\rangle \sum_{j=0}^{n-1} {}_{2n+1}C_{2j+1} \hat{P}_1^{2n-2j} \rho_d^1 \hat{P}_1^{2j+1} \sum_{k=0}^{m-1} {}_{2m}C_{2k+1} \hat{P}_2^{2m-2k-1} \rho_d^2 \hat{P}_2^{2k+1} \Big)$$

$$+ \sum_{n=0}^{\infty} \frac{(-ig)^{2n+1}}{(2n+1)!} \sum_{m=0}^{\infty} \frac{(-ig)^{2m+1}}{(2m+1)!} \Big( -\langle\psi_f|\hat{A}|\psi_i\rangle\langle\psi_i|\hat{B}|\psi_f\rangle \sum_{j=0}^{n} {}_{2n+1}C_{2j} \hat{P}_1^{2n+1-2j} \rho_d^1 \hat{P}_1^{2j} \sum_{k=0}^{m-1} {}_{2m+1}C_{2k+1} \hat{P}_2^{2m-2k} \rho_d^2 \hat{P}_2^{2k+1}$$

$$- \langle\psi_f|\hat{B}|\psi_i\rangle\langle\psi_i\hat{A}|\psi_f\rangle \sum_{j=0}^{n-1} {}_{2n+1}C_{2j+1} \hat{P}_1^{2n-2j} \rho_d^1 \hat{P}_1^{2j+1} \sum_{k=0}^{m} {}_{2m+1}C_{2k} \hat{P}_2^{2m+1-2k} \rho_d^2 \hat{P}_2^{2k} \Big) \Big]. \qquad (C19)$$

As in equation (C15), the expectation value of the joint pointer positions is obtained as

$$\langle \hat{X}_i \hat{X}_j \rangle_{\psi_f \perp \psi_i} = \frac{1}{2 \, Tr\rho'\Pi_f} \text{Re}[\langle\psi_f|\hat{A}|\psi_i\rangle\langle\psi_i|\hat{B}|\psi_f\rangle]. \qquad (C20)$$

**Appendix D: Local hidden-variable model describing our results**

TABLE I. Probability distribution of all possible outcomes of CHSH test.

| $\lambda$ | $A$ | $A'$ | $B$ | $B'$ | $P(\lambda_i)$ | $\lambda$ | $A$ | $A'$ | $B$ | $B'$ | $P(\lambda_i)$ |
|---|---|---|---|---|---|---|---|---|---|---|---|
| $\lambda_1$ | +1 | +1 | +1 | +1 | $P_1$ | $\lambda_9$ | −1 | +1 | +1 | +1 | $P_9$ |
| $\lambda_2$ | +1 | +1 | +1 | −1 | $P_2$ | $\lambda_{10}$ | −1 | +1 | +1 | −1 | $P_{10}$ |
| $\lambda_3$ | +1 | +1 | −1 | +1 | $P_3$ | $\lambda_{11}$ | −1 | +1 | −1 | +1 | $P_{11}$ |
| $\lambda_4$ | +1 | +1 | −1 | −1 | $P_4$ | $\lambda_{12}$ | −1 | +1 | −1 | −1 | $P_{12}$ |
| $\lambda_5$ | +1 | −1 | +1 | +1 | $P_5$ | $\lambda_{13}$ | −1 | −1 | +1 | +1 | $P_{13}$ |
| $\lambda_6$ | +1 | −1 | +1 | −1 | $P_6$ | $\lambda_{14}$ | −1 | −1 | +1 | −1 | $P_{14}$ |
| $\lambda_7$ | +1 | −1 | −1 | +1 | $P_7$ | $\lambda_{15}$ | −1 | −1 | −1 | +1 | $P_{15}$ |
| $\lambda_8$ | +1 | −1 | −1 | −1 | $P_8$ | $\lambda_{16}$ | −1 | −1 | −1 | −1 | $P_{16}$ |

Here, we present a hidden-variable model describing the classical results obtained from our DFM. To test the CHSH inequality, we need four binary observables: $A$, $A'$, $B$, and $B'$. Thus, 16 hidden variables, $\lambda_i$, and the corresponding probabilities, $P_i$, are required, as shown in Table I. Then, we find a set of constraints; normalization expressed as $\sum_i P_i = 1$, the averages of a single observable $\langle A \rangle_{\text{DFM}} = \langle A' \rangle_{\text{DFM}} = \langle B \rangle_{\text{DFM}} = \langle B' \rangle_{\text{DFM}} = 0$ and those of correlations $\langle AB \rangle_{\text{DFM}} = \langle A'B \rangle_{\text{DFM}} = \langle AB' \rangle_{\text{DFM}} = -\langle A'B' \rangle_{\text{DFM}} = -\sqrt{2}/4$. These constitute a system of nine linear equations of variable $P_i$ values, written as

$$\begin{pmatrix} +1 & +1 & +1 & +1 & +1 & +1 & +1 & +1 & +1 & +1 & +1 & +1 & +1 & +1 & +1 & +1 \\ +1 & +1 & +1 & +1 & +1 & +1 & +1 & +1 & -1 & -1 & -1 & -1 & -1 & -1 & -1 & -1 \\ +1 & +1 & +1 & +1 & -1 & -1 & -1 & -1 & +1 & +1 & +1 & +1 & -1 & -1 & -1 & -1 \\ +1 & +1 & -1 & -1 & +1 & +1 & -1 & -1 & +1 & +1 & -1 & -1 & +1 & +1 & -1 & -1 \\ +1 & -1 & +1 & -1 & +1 & -1 & +1 & -1 & +1 & -1 & +1 & -1 & +1 & -1 & +1 & -1 \\ +1 & +1 & -1 & -1 & +1 & +1 & -1 & -1 & -1 & -1 & +1 & +1 & -1 & -1 & +1 & +1 \\ +1 & +1 & -1 & -1 & -1 & -1 & +1 & +1 & +1 & +1 & -1 & -1 & -1 & -1 & +1 & +1 \\ +1 & -1 & +1 & -1 & +1 & -1 & +1 & -1 & -1 & +1 & -1 & +1 & -1 & +1 & -1 & +1 \\ +1 & -1 & +1 & -1 & -1 & +1 & -1 & +1 & +1 & -1 & +1 & -1 & -1 & +1 & -1 & +1 \end{pmatrix} \begin{pmatrix} P_1 \\ P_2 \\ P_3 \\ P_4 \\ P_5 \\ P_6 \\ P_7 \\ P_8 \\ P_9 \\ P_{10} \\ P_{11} \\ P_{12} \\ P_{13} \\ P_{14} \\ P_{15} \\ P_{16} \end{pmatrix} = \begin{pmatrix} 1 \\ 0 \\ 0 \\ 0 \\ 0 \\ 0 \\ -\sqrt{2}/4 \\ -\sqrt{2}/4 \\ -\sqrt{2}/4 \\ \sqrt{2}/4 \end{pmatrix}. \quad \text{(D1)}$$

Because the number of equations is smaller than that of unknowns, this is an indeterminate system. Thus, we have an infinite number of hidden-variable models describing our results.

**Appendix E: Leggett–Garg inequality via DFM**

If two postulates, the macroscopic realism and non-invasive measurability, are established, the Leggett–Garg inequality (LGI) is satisfied:

$$\langle I_{t_1} I_{t_2}\rangle + \langle I_{t_2} I_{t_3}\rangle - \langle I_{t_1} I_{t_3}\rangle \leq 1, \tag{E1}$$

where $I_{t_i}$ indicates a measurement outcome of the macroscopic observable $I$ at time $t_i$ ($t_1 < t_2 < t_3$), and these terms are dichotomic, i.e., $I_{t_i} = \pm 1$. Macroscopic realism implies that measurement of a macroscopic system reveals a well-defined pre-existing value. The postulate of non-invasive measurability states that, in principle, we can measure this value without disturbing the system. This inequality is regarded as a form of "quantumness," similar to the Bell inequality.

If we apply our DFM to the LGI, we trivially find that the LGI is satisfied. The expectation value of two observables obtained at two distinct times can be separated as

$$\langle \hat{I}_{t_i}\rangle_{\text{DFM}} \langle \hat{I}_{t_j}\rangle_{\text{DFM}} = \frac{\langle\psi_i|\hat{I}_{t_i}|\psi_i\rangle}{\langle\psi_i|\psi_i\rangle} \frac{\langle\psi_i|\hat{I}_{t_j}|\psi_i\rangle}{\langle\psi_i|\psi_i\rangle} = \langle \hat{I}_{t_i}\rangle \langle \hat{I}_{t_j}\rangle, \tag{E2}$$

because in the DFM scheme, the entire procedure is completed with projection measurement for post-selection. The measurement of $\hat{I}_{t_i}$ collapses the initial state $\psi_i$ to the initial state at $t_i$, and we subsequently perform our DFM on $\hat{I}_{t_j}$ with the same $\psi_i$ at $t_j$ ($t_i < t_j$).

**Appendix F: Simulation results corresponding to Figure 4**

Figure 7 shows the simulated detection probability distributions depending on the post-selected states. These simulations can be conducted using the equations given in Appendix C. However, this approach is cumbersome, because of the excessive number of terms in these expressions. Here, we show a direct calculation method for the device (pointer) state distributions after post-selection. The initial states of the system ($|\psi^-\rangle_{12}$) and two devices are separable, as follows:

$$|\Psi\rangle = |\psi^-\rangle_{12}|\psi_G(x_0,\sigma)\rangle_{d_1}|\psi_G(y_0,\sigma)\rangle_{d_2}, \tag{F1}$$

where $|\psi_G(c_0, \sigma)\rangle_{d_i}$ is the $i^{th}$ device state prepared as a Gaussian state with center $c_0$ and width $\sigma$. We assume that the initial center positions are zero and that the two device states have the same width. A measurement process is a translation operation $T$ to the device state depending on the measurement basis states. Therefore, the joint measurement has the form

$$J = [|1_A\rangle_1\langle 1_A|T_{x_1}(1) + |-1_A\rangle_1\langle -1_A|T_{x_1}(-1)] \otimes [|1_B\rangle_2\langle 1_B|T_{x_2}(1) + |-1_B\rangle_2\langle -1_B|T_{x_2}(-1)],\tag{F2}$$

where $|\pm 1_M\rangle$ is the eigenstate of the measurement $M$ with eigenvalue $\pm 1$. The degree of device translation corresponds to the eigenvalues. After post-selection to a final state $|\psi_f\rangle$, the wavefunction of the two devices has the form of $\langle x_1|\langle x_2|\langle \psi_f|J|\Psi\rangle$. Therefore, the

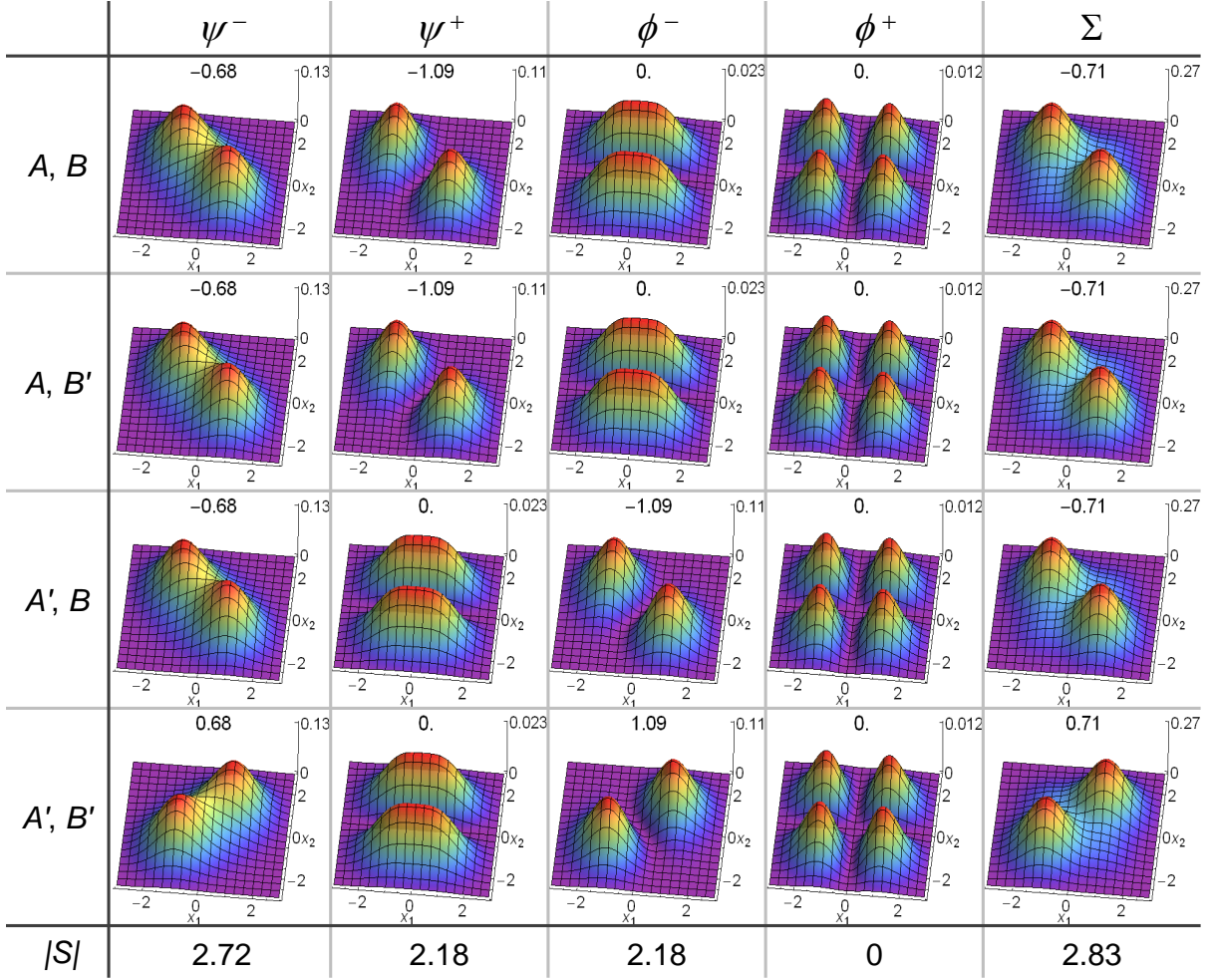

FIG. 7. Simulation results of detection probability distributions with identical experimental conditions to Figure 4.

two-dimensional detection probability distribution without normalization is

$$P_{\psi_f}(x_1, x_2) = |\langle x_1|\langle x_2|\langle \psi_f|J|\Psi\rangle|^2. \tag{F3}$$

The simulation results in Figure 7 were calculated using equation (F3) and the effective $\sigma$, which was experimentally measured and renormalized with an experimental translation distance $g$.

[39] The smaller $s$, the smaller the measurement probability for the orthogonal state. Therefore, the measurement time becomes longer to keep the measurement results meaningful (to reduce the uncertainty). In this experiment, because of the wavelength stability of the pump laser, the experiment for each $s$ had to be completed within two hours. Even though $s < 0.1$ can be realized, the uncertainty of the experimental result for $s \simeq 0.1$ within the time limit is too wide ($\Delta |S_{\phi^+}| \simeq 2$).